\documentstyle[aps,prb,twocolumn,epsf,floats]{revtex}
\begin{document}
\draft
\wideabs{
\title{
   Quasiparticle Interactions in Fractional Quantum Hall Systems:\\
   Justification of Different Hierarchy Schemes}
\author{
   Arkadiusz W\'ojs$^{ab*}$ and John J. Quinn$^{a*}$}
\address{
   (a) Department of Physics, University of Tennessee, 
   Knoxville, Tennessee 37996\\
   (b) Institute of Physics, Wroclaw University of Technology,
   50-370 Wroclaw, Poland}
\maketitle
\begin{abstract}
   The pseudopotentials describing the interactions of quasiparticles 
   in fractional quantum Hall (FQH) states are studied.
   Rules for the identification of incompressible quantum fluid ground 
   states are found, based upon the form of the pseudopotentials.
   States belonging to the Jain sequence $\nu=n(1+2pn)^{-1}$, where
   $n$ and $p$ are integers, appear to be the only incompressible 
   states in the thermodynamic limit, although other FQH hierarchy 
   states occur for finite size systems.
   This explains the success of the composite Fermion picture.
\end{abstract}
}

\section{Introduction}
The fractional quantum Hall (FQH) effect,\cite{tsui,laughlin-fqhe,prange,%
stormer} i.e. quantization of the Hall conductance of a two dimensional 
electron gas (2DEG) at certain densities in high magnetic fields, is a 
consequence of a finite excitation gaps that open above (incompressible) 
non-degenerate ground states (GS's) at certain fractional fillings of the 
lowest Landau level (LL), $\nu=1/3$, 2/5, etc.
Because of the LL degeneracy, it is clear that the those incompressible 
GS's must originate from electron-electron interactions.
This makes physics of the FQH effect very different from that underlying 
the integer quantum Hall (IQH) effect,\cite{klitzing,laughlin-iqhe} 
despite similar manifestation of the two in transport experiments.

While the origin of most prominent FQH states at $\nu=1/m$ ($m$ is an 
odd integer) has been explained by Laughlin,\cite{laughlin-fqhe} our 
understanding of why such states are also formed at some other filling 
factors (independent of the sample, density, etc.) or what conditions 
must in general be satisfied for an interacting system to exhibit the 
FQH effect at different fractions, is not yet completely satisfactory.
As an extension of Laughlin's idea, Haldane,\cite{haldane-hierarchy} 
and others\cite{laughlin-hierarchy,halperin-hierarchy,jain-hierarchy,%
sitko-hierarchy} proposed different, although equivalent,
\cite{read,chen,yang} hierarchy schemes, in which the elementary 
quasiparticle-like excitations of the Laughlin fluid form Laughlin like 
states of their own.
However, the original hierarchy approaches all share a major problem: 
they predict too many fractions and give no apparent connection between 
the stability of a given state and the hierarchy level at which it occurs.
It is known\cite{tsui,willet,mallet,sajoto,kukushkin} that the FQH effect 
does not occur at some simple fractions predicted at low levels of 
hierarchy, while explanation of some other, experimentally observed ones 
requires introducing many generations of excitations.

A different approach, introduced by Jain,\cite{jain} and developed 
by Lopez and Fradkin,\cite{lopez} and by Halperin, Lee, and Read
\cite{halperin-lee-read} involves the concept of composite Fermions 
(CF's).
As formally described by the mean field Chern-Simons (CS) transformation, 
the CF's are constructed by binding part of the external magnetic field 
$B$ to electrons in form of infinitely thin solenoids each carrying an 
even number $2p$ of flux quanta.
The resulting CF's move in an effective, reduced field $B^*$.
Smaller LL degeneracy of CF's leads to a lower effective filling factor
$\nu^*$ given by
\begin{equation}
   (\nu^*)^{-1}=\nu^{-1}-2p.
\label{eqjain}
\end{equation}
It was found that the sequence of fractional electron filling factors 
$\nu$ corresponding to integer CF fillings $\nu^*$ agrees with the 
values at which FQH states occur.
The qualitative success of the CF picture led to the conjecture that the 
CS charge-flux and Coulomb charge-charge interactions beyond the mean 
field largely cancel one another, and the system of strongly interacting 
electrons in a high magnetic field is converted to one of weakly 
interacting CF's in a lower field.
However, this cannot possibly be correct, since the CS energy ($\propto 
B$) and Coulomb energy ($\propto\sqrt{B}$) scale differently with the 
magnetic field $B$.
Also, higher LL's used in the procedure to obtain the wavefunctions are 
later eliminated by projection, which puts in doubt their role in the 
final result and most likely does not allow for interpretation in terms 
of an additional, effective magnetic field.

In this paper we attempt to justify Haldane's hierarchy picture in terms 
of the behavior of the interaction between quasiparticles (QP's) which he 
considered of importance but totally unknown.
We derive simple rules for identifying fractions that do or do not 
correspond to incompressible FQH states by analyzing pseudopotentials
\cite{haldane-pseudo,wojs-parent1,wojs-parent2,wojs-parent3} of the 
QP interactions on successive levels of hierarchy.
Based on the connection between the form of pseudopotential (pair energy 
vs. pair angular momentum) and the nature of low lying states, we explain 
why QP's do not form Laughlin states at all $1/m$ fillings.
Our results validate understanding of experimentally observed FQH states 
in intuitive terms of the hierarchy of Laughlin excitations, established 
in Refs.~\onlinecite{haldane-hierarchy,laughlin-hierarchy,%
halperin-hierarchy,jain-hierarchy,sitko-hierarchy}.
We also show that (in large systems) valid incompressible FQH states 
obtained in this hierarchy picture are equivalent to Jain states, despite 
different underlying physics used to justify the two approaches.
This explains the success of the CF picture when applied to FQH systems
better than assumed cancellation between CS and Coulomb interactions.
It also defines limitations of the CF picture when applied to systems 
with general interactions (e.g., recently studied FQH systems of charged
excitonic complexes formed in an electron-hole gas in a high magnetic 
field\cite{wojs-xminus1,wojs-xminus2,wojs-2comp}).
The discussion is illustrated with results of exact numerical 
diagonalization calculations on a Haldane sphere\cite{haldane-hierarchy}
of up to twelve electrons at $\nu\approx1/3$, (hamiltonians with 
dimensions of up to $2\times10^6$) carried up using a modified 
Lanczos\cite{lanczos} algorithm.

\section{Haldane sphere}
Because in the absence of electron-electron interactions all many body 
states in the lowest LL ($\nu\le1$) are degenerate, those interactions
cannot be treated perturbatively.
Therefore, numerical diagonalization techniques are commonly used to 
study FQH systems, which, however, limit their size to a finite (small) 
number of electrons.
In order to model a finite density 2DEG, motion of $N$ electrons must be 
restricted to a finite area.
This can be accomplished by imposing lateral confinement,\cite{laughlin-conf} 
periodic boundary conditions (PBC),\cite{haldane-pbc} or confining electrons 
on a closed surface (Haldane sphere).\cite{haldane-hierarchy}

The last approach is particularly useful as it naturally avoids edge effects 
and preserves full 2D translational symmetry of a planar 2DEG in form of the 
rotational symmetry of a sphere.
A pair of good quantum numbers on a plane, the center of mass (CM) and 
relative momenta, correspond to the total angular momentum $L$ and its 
projection $L_z$ on a sphere.\cite{wojs-geometry} 
The degeneracies associated with CM excitations on a plane correspond 
to those of $L$ multiplets on a sphere, and the non-degenerate GS's of 
a planar 2DEG have $L=0$ in their spherical models.

The magnetic field $B$ perpendicular to the surface of the Haldane sphere 
of radius $R$ is an isotropic radial field produced by a magnetic monopole 
placed at the origin.
The monopole strength $2S$, defined as the number of flux quanta piercing 
the sphere, is an integer, as required by Dirac's condition.\cite{dirac} 
In extrapolation of finite size results to the thermodynamic limit, the 
magnetic length $\lambda=R/\sqrt{S}$ is used as the length scale.
The single particle states on the Haldane sphere are labeled by angular 
momentum $l\ge S$ and its projection $l_z$, and are called monopole 
harmonics.\cite{haldane-hierarchy,wu,fano}
The energies form $(2l+1)$-fold degenerate angular momentum shells,
or LL's, labeled by $n=l-S$ and separated by cyclotron gaps.
For the FQH states at $\nu<1$, only the lowest ($n=0$) spin polarized 
angular momentum shell of $l=S$ need be considered.
The $N$ electron Laughlin $\nu=1/m$ states in a $2l+1$ degenerate shell 
occur at $2l=m(N-1)$.

\section{Hierarchy of Laughlin excitations}
The Haldane's hierarchy of FQH states is constructed in the following way.
At certain filling factors, $\nu=1/m$ ($m$ is an odd integer), electrons 
form Laughlin incompressible states.
At $\nu$ slightly different from a Laughlin $1/m$ filling, the low lying 
states must contain a number of quasiparticles (QP's) --- quasielectrons 
(QE's) at $\nu>1/m$ or quasiholes (QH's) at $\nu<1/m$ --- in the Laughlin 
$\nu=1/m$ state.
States involving more than the minimum number of QP's required by the 
difference between $\nu$ and $1/m$ contain additional QE-QH pairs
\cite{delta} and are separated from the lowest band by a gap $\Delta$.
The QP's in the underlying (parent) Laughlin state have certain single 
particle energy $\varepsilon_{\rm QP}$, statistics, number of available 
single particle states $g_{\rm QP}$ (in analogy to LL degeneracy for 
electrons), and interact with one another.
Provided their interaction is small compared to $\Delta$ and 
$\varepsilon_{\rm QP}$, QP's can, in principle, form Laughlin 
incompressible (daughter) states of their own.
If $\nu$ is not exactly equal to the value at which QP's would form a 
daughter Laughlin state, low lying QP states will contain their own 
QP-like excitations in the nearest daughter Laughlin state, which can 
in turn form an incompressible granddaughter Laughlin state, etc.

Since the QE or QH statistics enters the hierarchy of fractions only 
through the counting of many body states, different statistics give 
equivalent results as long as the single particle degeneracies $g_{\rm 
QE}$ and $g_{\rm QH}$ are chosen correctly.
The mean field CS transformation can be used to formally convert Bosons 
into Fermions by attaching one flux quantum to each Boson.
\cite{wilczek,canright,xie}
For example, in the spherical geometry, where $g_{\rm QP}=2l_{\rm QP}+1$ 
is related to QP angular momentum $l_{\rm QP}$, systems of $N$ Bosons each 
with angular momentum $k$ and $N$ Fermions each with angular momentum $l$ 
are equivalent and contain the same $L$ multiplets if
\begin{equation}
   l=k+(N-1)/2.
\label{eqbf}
\end{equation}
As shown by Haldane\cite{haldane-hierarchy} (and can be understood from 
a simple picture of excitations created between electrons on a line), 
QE's and QH's in an $N$ electron Laughlin state can be viewed as Bosons 
with $k_{\rm QE}=k_{\rm QH}=N/2$.
However, this combination of statistics and degeneracy gives proper 
counting of many body states only if an additional hard core is introduced 
that forbids two QE's to be in a pair state with $L=N$.\cite{he}
Such hard core can be accounted for by a mean field Fermion-to-Fermion 
CS transformation that replaces $k_{\rm QE}$ by $k_{\rm QE}^*=k_{\rm QE}
-(N_{\rm QE}-1)$,\cite{wojs-parent1,wojs-parent2,wojs-parent3} where 
$N_{\rm QP}$ is the QP number.

In order to stress the connection with Jain's CF picture and the recently
proposed\cite{sitko-hierarchy} hierarchy of CF excitations, we use here 
a fermionic description of QP's.
The appropriate QP angular momenta obtained from Eq.~(\ref{eqbf}) are
\begin{equation}
   l_{\rm QP}={N\pm(n_{\rm QP}-1)\over2},
\label{eqdeg}
\end{equation}
with $(-)$ for $l_{\rm QE}$ and $(+)$ for $l_{\rm QH}$.
Note that $l_{\rm QE}$ and $l_{\rm QH}$ given by Eq.~(\ref{eqdeg}) are 
equal to angular momenta of holes in highest filled and particles in 
lowest empty CF LL's used in the CF picture.

\begin{figure}[t]
\epsfxsize=3.35in
\epsffile{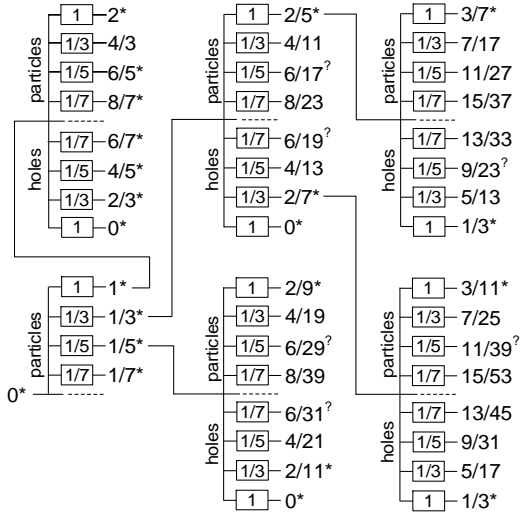}
\caption{
   Diagram of filling factors obtained in the hierarchy of Laughlin
   excitations. Lines connect parent and daughter states with 
   quasiparticle filling factors shown in boxes. Asterisks -- 
   incompressible states, question marks -- hierarchy ground states 
   observed only in finite systems, other fractions -- compressible 
   states.
}
\label{fig1}
\end{figure}

The expression for the filling factor $\nu$ of a daughter state is 
very similar to Eq.~(\ref{eqjain}) and reads\cite{sitko-hierarchy}
\begin{equation}
   \nu^{-1}=2p+(1\pm\nu_{\rm QP})^{-1},
\label{eqcont}
\end{equation}
where $\nu_{\rm QP}$ is the filling factor of QE's $(+)$ or QH's $(-)$ 
in Laughlin $(2p+1)^{-1}$ parent state.
Iteration of Eq.~(\ref{eqcont}) with $\nu_{\rm QP}$ substituted by 
$\nu$ gives a continuous fraction, terminated when an incompressible 
state (without QP's, $\nu_{\rm QP}=0$) is reached.
E.g., for the state containing $\nu=1/5$ filling of QH's in the $\nu=1$ 
filling of QE's in the Laughlin $\nu=1/3$ state of electrons, this 
procedure gives:
\begin{equation}
   \nu={\displaystyle1\over\displaystyle2\cdot1+
       {\displaystyle1\over\displaystyle1+
       {\displaystyle1\over\displaystyle2\cdot0+
       {\displaystyle1\over\displaystyle1-
       {\displaystyle1\over\displaystyle2\cdot2+
       {\displaystyle1\over\displaystyle1+0}}}}}}={9\over23}.
\end{equation}
In Fig.~\ref{fig1} we display more filling factors obtained in the 
hierarchy scheme.
The lines connect parent states with their daughter states obtained
for QP fillings $\nu_{\rm QP}$ shown in boxes.
Note that holes created in a parent state of holes are particles in 
the grandparent state (in a sense that their presence increases overall 
$\nu$, i.e. their number increases when the magnetic field is decreased).
E.g., holes created in $\nu=1/3$ filling of holes in $\nu=1/3$ Laughlin 
state of electrons will be referred to as QE's in the $\nu=2/7$ state.

As first stated by Haldane,\cite{haldane-hierarchy} whether a given 
fraction obtained in the hierarchy scheme corresponds to a stable 
incompressible GS depends on the stability of the parent state and 
on the interaction between QP's in the daughter state.
As will be shown in the following sections, these criteria eliminate 
most of all odd denominator fractions that can be constructed by an
iteration of Eq.~(\ref{eqcont}).
The relatively small number of possible candidates left include the 
Jain sequence obtained from Eq.~(\ref{eqjain}) for integer $\nu^*$.
These states (and their electron-hole conjugates) have been marked 
with asterisks in Fig.~\ref{fig1}.
As will be shown in the following sections, these are all the
incompressible states predicted by the ``correct'' hierarchy picture 
in the thermodynamic limit, and all of them have been confirmed 
experimentally.\cite{tsui,willet,mallet,sajoto,kukushkin} 
The fractions with a question mark in Fig.~\ref{fig1} are most likely 
compressible in the thermodynamic limit, but valid non-degenerate ($L=0$) 
hierarchy GS's can occur at these fillings in finite systems.
Those finite size valid hierarchy GS's (Laughlin states of Laughlin QP's) 
should be distinguished from other $L=0$ GS's that can occur at different 
combinations of $N$ and $2S$ (e.g., for $N=12$ and $2S=29$ in Fig.~3c
of Ref.~\onlinecite{sitko-hierarchy}), but do not have Laughlin like QP 
correlations and thus cannot be associated with a filling factor.
At the remaining filling factors, unmarked in Fig.~\ref{fig1}, the 
hierarchy picture fails as expected and the system is compressible.

\section{Pseudopotentials}
The two body interaction of identical particles in an angular momentum 
shell of degeneracy $2l+1$ can be written in terms of the pseudopotential 
\cite{haldane-pseudo} $V({\cal R})$, i.e. the pair interaction energy 
$V$ as a function of relative pair angular momentum 
\begin{equation}
   {\cal R}=2l-L,
\end{equation}
where $L$ is the total angular momentum of the pair.\cite{ram}
${\cal R}$ is an odd integer and increases with increasing average 
separation.\cite{wojs-parent3}
Plotting $V$ as a function of ${\cal R}$ rather than of $L$ allows 
for meaningful comparison of pseudopotentials in shells of different 
degeneracy; for $l\rightarrow\infty$, the pseudopotentials calculated 
on the sphere converge to the pseudopotential on a plane (on a plane,
${\cal R}$ is defined\cite{wojs-geometry} as the usual relative angular 
momentum).

Whether a system of interacting Fermions will form a Laughlin state at 
the $1/m$ filling of their angular momentum shell depends on the short 
range of repulsive interaction.
Precisely, the Laughlin $\nu=1/m$ incompressible state (in which ${\cal 
R}\ge m$ for all pairs) is formed if,\cite{wojs-parent3} in the vicinity 
of ${\cal R}=m$, the interaction pseudopotential increases more quickly 
than linearly as a function of $L(L+1)$, i.e. more quickly than that of 
harmonic repulsion, $V_{\rm H}(L)=\alpha+\beta L(L+1)$, where $\alpha$ 
and $\beta$ are constants.
More generally, if this condition is satisfied in the vicinity of 
${\cal R}=2p+1$ for certain $p$, the total many body Hilbert space 
${\cal H}$ contains an (approximate) eigensubspace\cite{wojs-parent3} 
${\cal H}_p$ holding states with ${\cal R}\ge2p+1$ for all pairs (i.e. 
avoiding $p$ pair states of largest repulsion).
A corresponding low energy band occurs in the spectrum, separated from 
higher states by a gap associated with $V({\cal R})$.
At Laughlin fillings of $\nu=(2p+1)^{-1}$, the subspace ${\cal H}_p$ 
contains a single non-degenerate ($L=0$) multiplet with ${\cal R}\ge2p+1$,
and the lowest band consists of the Laughlin GS.

The mathematical formalism derived to quantitatively treat the ability of 
electrons to avoid certain pair states involves the concept of fractional 
(grand)parentage,\cite{shalit,cowan} well known in atomic and nuclear 
physics and used recently\cite{wojs-parent1,wojs-parent2,wojs-parent3} 
to describe FQH systems.
It is worth noting that avoiding highest energy states of three or more 
particles was recently proposed\cite{ginocchio} to explain incompressible 
GS's at other fillings than Laughlin's $\nu=1/m$.

\begin{figure}[t]
\epsfxsize=3.35in
\epsffile{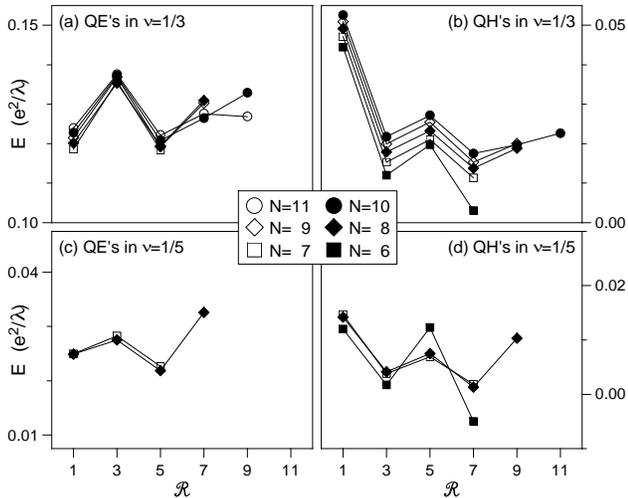}
\caption{
   Energies $E=2\varepsilon+V({\cal R})$ of a pair of quasielectrons
   (left) and quasiholes (right) in Laughlin $\nu=1/3$ (top) and 
   $\nu=1/5$ (bottom) states, as a function of relative pair angular 
   momentum ${\cal R}$, obtained in diagonalization of $N$ electrons. 
}
\label{fig2}
\end{figure}

The electron (Coulomb) pseudopotential in the lowest LL $V_{\rm e}
({\cal R})$ satisfies\cite{wojs-parent1,wojs-parent2,wojs-parent3} 
the ``short range'' criterion (i.e. increases more quickly than 
$V_{\rm H}$ as a function of $L$) in entire range of ${\cal R}$, which 
is the reason for incompressibility of principal Laughlin $\nu=1/m$ 
states.
However, this does not generally hold for the QP pseudopotentials
\cite{beran,yi} on higher levels of hierarchy.
We have obtained some of those pseudopotentials for different values of
$l_{\rm QP}$ by numerical diagonalization of appropriate many electron 
hamiltonians on the Haldane sphere and identification of the lowest bands 
in obtained energy spectra.
The total many electron energies within those bands contain the energy 
$E_0$ of the parent state, single particle energies $2\varepsilon_{\rm 
QP}$ of the pair of appropriate QP's, and the QP-QP interaction energy 
$V_{\rm QP}(L)$.
In Fig.~\ref{fig2} we show the results for QE's and QH's in Laughlin 
$\nu=1/3$ (data for $N\le8$ was published before in Ref.~\onlinecite{yi}) 
and $\nu=1/5$ states.
The plotted energy $E({\cal R})=2\varepsilon_{\rm QP}+V_{\rm QP}(L)$ is 
given in units of $e^2/\lambda$ where $\lambda$ is the magnetic length 
in the parent state.
Different symbols mark pseudopotentials obtained in diagonalization of
$N$ electron systems with different $N$ and thus with different $l_{\rm 
QP}$, see Eq.~(\ref{eqdeg}).
Clearly, the QE and QH pseudopotentials are quite different and neither
one decreases monotonically with increasing ${\cal R}$.
On the other hand, the corresponding pseudopotentials in $\nu=1/3$ 
and 1/5 states look similar, only the energy scale is different.
The convergence of energies at small ${\cal R}$ obtained for larger $N$ 
suggests that the maxima at ${\cal R}=3$ for QE's and at ${\cal R}=1$ 
and 5 for QH's, as well as the minima at ${\cal R}=1$ and 5 for QE's 
and at ${\cal R}=3$ and 7 for QH's, persist in the limit of large $N$ 
(i.e. for an infinite system on a plane).
Consequently, the only incompressible daughter states of Laughlin 
$\nu=1/3$ and 1/5 states are those with $\nu_{\rm QE}=1$ or $\nu_{\rm 
QH}=1/3$ (asterisks in Fig.~\ref{fig1}) and (maybe) $\nu_{\rm QE}=1/5$ 
and $\nu_{\rm QH}=1/7$ (question marks in Fig.~\ref{fig1}).
It is also clear that no incompressible daughter states will form at 
e.g. $\nu=4/11$ or 4/13.

Let us note that the incompressibility of daughter states with completely 
filled QE shell (e.g., at $\nu=2/5$ or 2/9) does not require any special 
form of the QE-QE interaction, except that it must be weaker than the 
single particle energies $\varepsilon_{\rm QE}$ and $\varepsilon_{\rm QH}$ 
responsible for the gap.
In this sense, the FQH effect at Jain filling $\nu=2/5$ can be viewed as 
an IQH effect of QE's in the Laughlin $\nu=1/3$ state, except that the 
degenerate single particle shell available to QE's is due to special form 
of elementary excitations of the parent Laughlin state rather than due to 
an effective magnetic field.
Similarly, the excitation gap at the $\nu=2/5$ filling is not a cyclotron
gap but the energy needed to create a QE-QH pair like excitation in the
filled shell of Laughlin QE's.
On the other hand, the FQH effect at $\nu=2/7$ is a fractional effect also 
on the level of QH's in the parent $\nu=1/3$ state, and its excitation gap 
is governed by $V_{\rm QH}(1)$, the largest pseudopotential parameter 
for QH-QH interaction. 

The electron pseudopotential $V_{\rm e}({\cal R})$ is not strictly a short 
range one (for which $V(1)\gg V(3)\gg\dots$) and the associated hidden 
symmetry responsible for occurrence of eigensubspaces ${\cal H}_p$ and 
incompressible Laughlin states is only approximate.
Actually, a fairly small reduction of $V_{\rm e}(1)$ compared to 
$V_{\rm e}(3)$ that can be achieved in a wide quantum well leads to 
a break down of Laughlin $\nu=1/3$ state.\cite{shayegan}
While the hidden symmetry makes the low lying states near Laughlin 
fillings virtually insensitive to the details of $V_{\rm e}({\cal R})$ 
as long as it has short range (i.e. increases more quickly than $V_{\rm 
H}$ as a function of $L$), it is interesting to ask to what extent the
form of $V_{\rm e}({\cal R})$ affects the QP pseudopotentials, and thus 
the incompressibility of related daughter states.
We have compared the pseudopotentials in Fig.~\ref{fig2} with the ones 
obtained for ``exponential'' interaction, $V_{\rm exp}({\cal R}+2)=V_{\rm 
exp}({\cal R})/10$, and for ``selective'' interaction, $V_{\rm sel}({\cal 
R}<m)=\infty$, $V_{\rm sel}(m)=1$, and $V_{\rm sel}({\cal R}>m)=0$, and 
found that all the features in Fig.~\ref{fig2} remain unchanged.
This means that the short range character of interaction between particles 
in the parent state does not imply the same for interaction between 
Laughlin QP's in the daughter state.
This observation, essential for understanding why incompressible states 
do not occur at all odd denominator fractions, might appear somewhat 
surprising since the QP's are (fractionally) charged objects and hence 
their interaction has similar nature to that of electrons.
However, it must be kept in mind that it is the combination of interaction 
potential $V(r)$ and available single particle Hilbert space that gives 
pseudopotential $V({\cal R})$ and, in turn, determines if Laughlin states 
occur (e.g., for the same Coulomb potential $V(r)=e^2/r$, Laughlin $\nu=
1/3$ state occurs in the lowest LL but does not occur in the excited 
ones\cite{wojs-parent3,macdonald}).

\begin{figure}[t]
\epsfxsize=3.35in
\epsffile{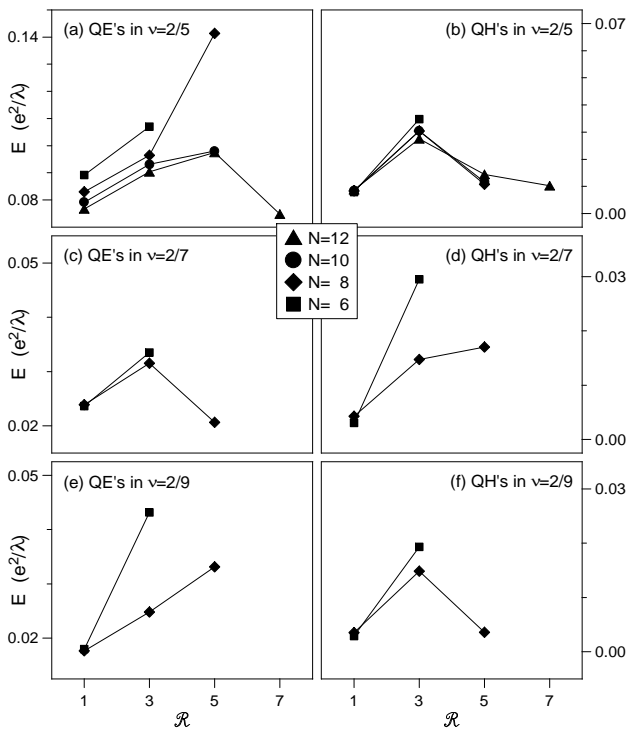}
\caption{
   Energies $E=2\varepsilon+V({\cal R})$ of a pair of quasielectrons
   (left) and quasiholes (right) in the second level hierarchy states 
   at $\nu=2/5$ (top), $\nu=2/7$ (center), and $\nu=2/9$ (bottom),
   as a function of relative pair angular momentum ${\cal R}$, obtained 
   in diagonalization of $N$ electron systems.
}
\label{fig3}
\end{figure}

Once the Laughlin like states on a given hierarchy level are found, the 
QP pseudopotentials must be calculated for these states to determine if 
they in turn can have any incompressible daughter states.
As an example of this procedure, in Fig.~\ref{fig3} we present a few 
pseudopotentials calculated for the $\nu=2/5$, 2/7, and 2/9 parent states.
As in Fig.~\ref{fig2}, energy $E({\cal R})=2\varepsilon_{\rm QP}+V_{\rm 
QP}(L)$ is given in the units of $e^2/\lambda$, with $\lambda$ appropriate 
for the parent state, and $N$ is the number of electrons in the system 
that was diagonalized to obtain a particular pseudopotential.

The pseudopotentials plotted in Fig.~\ref{fig3} show two types of behavior 
at small ${\cal R}$.
The ones in Fig.~\ref{fig3}bcf have a maximum at ${\cal R}=3$, similarly 
as those in Fig.~\ref{fig2}ac, while the ones in Fig.~\ref{fig3}ade 
increase when ${\cal R}$ increases between 1 and 5.
Similar behavior of different pseudopotentials is a consequence of the 
particle-hole symmetry between QE's of a parent state and QH's of its 
daughter state with filled QE shell ($\nu_{\rm QE}=1$, see Fig.~\ref{fig1}).
E.g., vacancies in an almost completely filled shell of QE's in the 
$\nu=1/3$ state are QH's of the $\nu=2/5$ state; vacancies in an almost 
completely filled shell of QE's in the $\nu=2/5$ state are QH's of the 
$\nu=3/7$ state, etc.
The relation between QH's in the parent state and QE's in the daughter 
state with $\nu_{\rm QH}=1$ is equivalent, as the latter is simply the 
grandparent state.
The particle-hole symmetry discussed above is only approximate because 
the single particle gaps are not infinitely large compared to single
particle gaps $\Delta$ and $\varepsilon$, and the QP number is not 
strictly conserved.
However, the appropriate pseudopotentials are to a good approximation
equal, and e.g. by comparing our data for QE's in $\nu=1/3$ and QH's 
in $\nu=2/5$ we were able to extract energies $\varepsilon_{\rm QH}$ of
a single QH in the $\nu=2/5$ state as a function of the system size $N$.
The linear extrapolation to $1/N\rightarrow0$ gives $\varepsilon_{\rm QH}
=0.0098$~$e^2/\lambda$, and the limiting value of the ``proper'' QH energy
\cite{fano,haldane-proper} (including additional fractional charge $e/5$ 
in the background) is $\tilde{\varepsilon}_{\rm QH}=0.0123$~$e^2/\lambda$.

It is apparent from Fig.~\ref{fig3} that the incompressible daughter 
states derived from $\nu=2/5$, 2/7, or 2/9 must either have a completely 
filled QE shell (Jain fractions $\nu=3/7$, 3/11, and 3/13, respectively), 
or (possibly) the 1/5 filling of QH's ($\nu=9/23$, 9/31, and 9/41,
respectively).
Incompressible daughter states at any other fractions, including the 1/3 
filling of QE's or QH's (giving such fractions as $\nu=5/13$ or 7/17, see
Fig.~\ref{fig1}) do not occur.

\section{Numerical tests for finite systems}
\begin{figure}[t]
\epsfxsize=3.35in
\epsffile{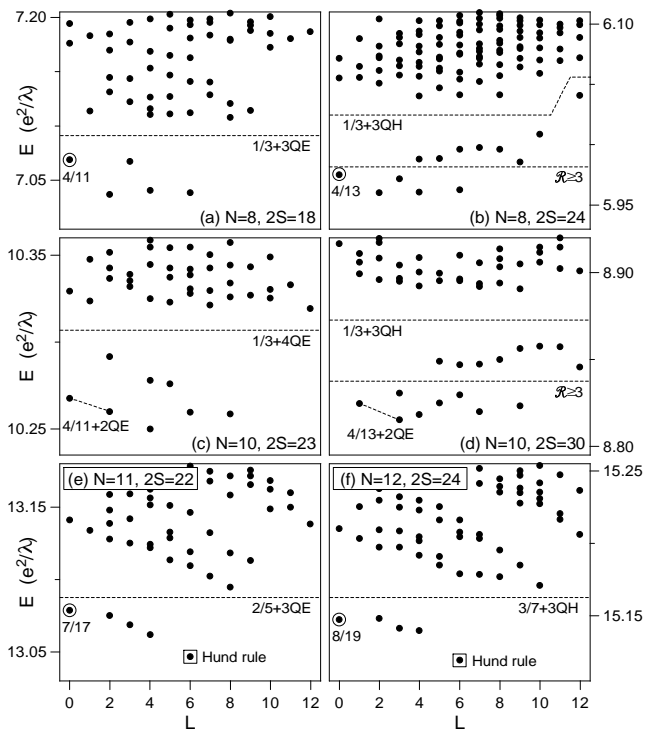}
\caption{
   Few electron energies $E$ as a function of total angular momentum 
   $L$, obtained in exact diagonalization in terms of individual 
   electrons interacting through the Coulomb pseudopotential. Dashed 
   lines mark gaps separating energy bands.
}
\label{fig4}
\end{figure}

In order to test the predictions of low lying states in terms of Laughlin 
QP's interacting through appropriate pseudopotentials, we have calculated 
numerically exact energy spectra of up to twelve electrons on the Haldane 
sphere at different values of the monopole strength $2S$, i.e., different 
filling factors.
As demonstrated on the examples presented in Fig.~\ref{fig4}, the results 
(in all cases we looked at) can be very well understood in terms of QP-QP 
interaction.
Fig.~\ref{fig4}a shows the spectrum of eight electrons at $2S=18$.
The low lying band contains states of three QE's in Laughlin $\nu=1/3$ 
state, each with $l_{\rm QE}=3$ ($L=0$, 2, 3, 4, and 6).
As marked with a dashed line, there is a gap separating the low energy
band from higher states.
Due to the QE-QE interaction, the lowest band is not degenerate and has
certain width.
Because this width is small compared to the gap to higher states (i.e. to
energy $\Delta$ to create additional QE-QH pairs), the three QE's interact 
with one another in the presence of a rigid background (Laughlin fluid at 
$\nu=1/3$) and the low lying states are determined by the pseudopotential 
$V_{\rm QE}({\cal R})$ obtained for the same $l_{\rm QE}=3$ ($N=7$ in 
Fig.~\ref{fig2}a).
If $V_{\rm QE}({\cal R})$ had short range, the multiplet at $L=0$ would 
be an incompressible GS corresponding to the Laughlin $\nu_{\rm QE}=1/3$ 
state and total electron filling factor of $\nu=4/11$.
However, as discussed in the previous section, $V_{\rm QE}({\cal R})$ has 
a minimum at ${\cal R}=1$ and the system is compressible (in this small 
system, the Laughlin $\nu_{\rm QE}=1/3$ state is an eigenstate as it is 
the only state of three QE's in the $L=0$ subspace; in larger systems, 
it will mix with other $L=0$ states and fall into the continuum).

A similar spectrum is displayed in Fig.~\ref{fig4}b.
Here, the low lying states contain three QH's in the Laughlin $\nu=1/3$ 
state, each with $l_{\rm QH}=5$.
As expected from the discussion of $V_{\rm QH}$, the states with ${\cal 
R}\ge3$ ($L=0$, 2, 3, 4, and 6) have lowest energy within this band, but
the Laughlin $\nu_{\rm QH}=1/5$ state with ${\cal R}\ge5$ in not the GS 
and the system at $\nu=4/13$ is compressible.

Since the GS's at $\nu=4/11$ or 4/13 are not valid parent states 
(QP Laughlin states), the analysis of states at $\nu$ near 4/11 or 4/13 
in terms of their daughter QP's is not possible.
This is demonstrated in Fig.~\ref{fig4}cd, where the states that would 
contain two QE's in the $\nu=4/11$ each with $l_{\rm QE}=3/2$ ($L=0$ and 
2) and two QE's in the $\nu=4/13$ each with $l_{\rm QE}=2$ ($L=1$ and 3), 
do not have lowest energy.

The low lying states in each of the spectra in Fig.~\ref{fig4}ef contain 
three QP's each with $l_{\rm QP}=3$ in a valid incompressible higher level 
hierarchy state ($L=0$, 2, 3, 4, and 6).
The interaction of QE's in the $\nu=2/5$ state and of QH's in the $\nu=3/7$ 
state is similar as those particles are connected through the particle-hole 
symmetry.
Also, since the appropriate pseudopotential in Fig.~\ref{fig3}a (for 
$N=10$) increases with increasing ${\cal R}$, we expect the standard atomic 
Hund rule to hold, i.e. the lowest energy state should have maximum allowed 
$L$ within the lowest energy band of three QP's.
Indeed, the $L=6$ state is the GS of both systems, while the states at 
$L=0$ predicted by the hierarchy picture ($\nu=7/17$ and 8/19) have higher 
energy.

Numerical tests of other fractions become difficult due to increasing
size of the system.
However, one can diagonalize the hamiltonian of interacting QP's in 
a given parent states using their known single particle energies 
$\varepsilon_{\rm QP}$ (the values for QP's in Laughlin $\nu=1/3$ and 
1/5 states can be found in Refs.~\onlinecite{wojs-parent3,fano})
and pseudopotentials $V_{\rm QP}({\cal R})$, and obtain approximate 
lowest energy levels of an underlying (larger) electron system (with 
respect to the energy $E_0$ of the parent incompressible state).
\cite{sitko-fermi}
The error made in such approximate calculation is due to neglected 
scattering processes involving other objects than the specified QP's 
(e.g. polarization of the parent state through creation of additional 
QE-QH pairs, etc.).
At least for states with largest $\varepsilon_{\rm QP}$ (compared to the 
strength of QP interactions), this error is expected to be small, which 
validates the tests of incompressibility of their daughter states.

\begin{figure}[t]
\epsfxsize=3.35in
\epsffile{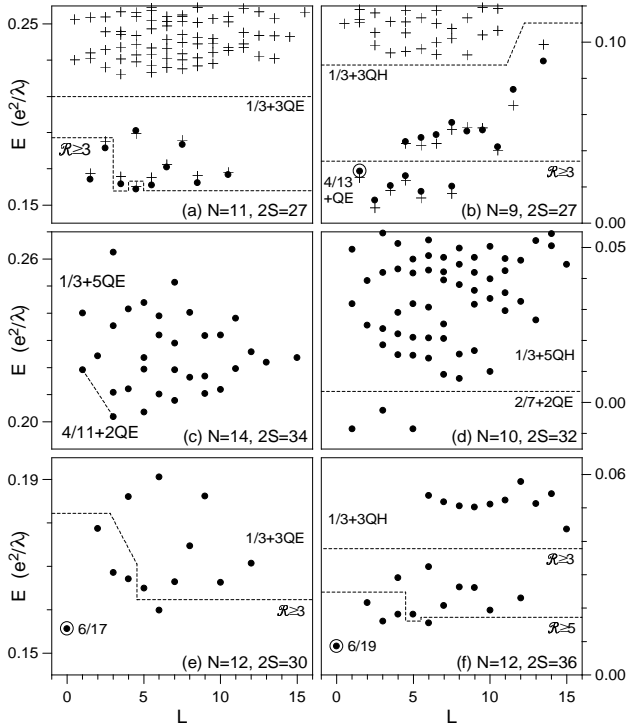}
\caption{
   Few electron energies $E$ measured from the energy of the parent 
   $\nu=1/3$ state as a function of total angular momentum $L$, 
   obtained in exact diagonalization in terms of Laughlin quasiparticles 
   (left) or quasiholes (right) of the $\nu=1/3$ state interacting 
   through appropriate pseudopotentials plotted in Fig.~\protect\ref{fig2}. 
   Dashed lines mark energy bands predicted by the hierarchy picture 
   assuming short range of all involved quasiparticle interactions. 
   Pluses in frames (a) and (b) mark exact energies obtained by 
   diagonalizing electron-electron interaction, as in 
   Fig.~\protect\ref{fig4}.
}
\label{fig5}
\end{figure}

The results of such tests for a few different systems are shown in 
Fig.~\ref{fig5}.
In all frames, the energy is measured from the energy of the parent 
($\nu=1/3$) state and $\lambda$ is the magnetic length in the parent 
state.
In Fig.~\ref{fig5}ab the approximate spectra (full dots) are overlaid 
with exact energies (pluses) obtained by diagonalization of the full 
electron hamiltonian.
Clearly, both for interacting QE's and QH's, the approximate calculation 
gives the lowest lying states with negligible error (most of which is 
due to different magnetic lengths in the daughter and parent states and 
can be corrected; error due to neglected scattering processes is hardly 
visible).
This agreement proves that the low lying states in FQH systems indeed 
contain QP's characterized by certain level degeneracy and interaction 
(weakly dependent on the QP number), and validates use of the approximate
calculation for larger systems.
The data in Fig.~\ref{fig5}a is by itself another example showing that 
QE's in Laughlin $\nu=1/3$ state do not form a separate band of states
with ${\cal R}\ge3$ (these would be states with $L=3/2$, 5/2, and 9/2).
Similarly, Fig.~\ref{fig5}b shows that QH's form a band with ${\cal R}
\ge3$ ($L=3/2$, 5/2, 7/2, 9/2, 11/2, and 15/2), but do not form a band
with ${\cal R}\ge5$ (here, one multiplet with $L=3/2$ that would correspond 
to a QE in the $\nu=4/13$ state).

Fig.~\ref{fig5}cdef shows approximate spectra of larger systems.
Fig.~\ref{fig5}c shows no band corresponding to two QE's in the $\nu=4/11$ 
state of fourteen electrons (these would be states at $L=1$ and 3), 
Fig.~\ref{fig5}d shows a band corresponding to two QE's in the $\nu=2/7$ 
state of ten electrons ($L=1$, 3, and 5; this is the QE pseudopotential 
like those for $N=6$ and 8 in Fig.~\ref{fig3}c),
Fig.~\ref{fig5}e shows no band of (Laughlin $\nu=1/3$) QE states with 
${\cal R}\ge3$ (these would be states at $L=0$, 2, 3, 4, and 6), and an 
incompressible $\nu=6/17$ state of twelve electrons corresponding to 
$\nu_{\rm QE}=1/5$, and 
Fig.~\ref{fig5}f shows a band of (Laughlin $\nu=1/3$) QH states with 
${\cal R}\ge3$, no band with ${\cal R}\ge5$ (these would be states at 
$L=0$, 2, 3, 4, and 6), and an incompressible $\nu=6/19$ state of twelve 
electrons corresponding to $\nu_{\rm QE}=1/7$.

The $\nu=6/17$ and 6/19 states of twelve electrons in Fig.~\ref{fig5}ef
are the only non-Jain hierarchy states we have tested numerically which 
are predicted to be (incompressible) $L=0$ GS's based on pseudopotentials 
in Figs.~\ref{fig2} and \ref{fig3} and which contain at least three QP's.
However, because of the particle-hole symmetry and the similarity of 
different QP pseudopotentials (see Figs.~\ref{fig2} and \ref{fig3}), 
the numerical evidence for valid $L=0$ hierarchy ground states at 
$\nu=6/17$ and 6/19 (in a twelve electron system) suggests stability 
of some other hierarchy states, e.g. at $\nu=6/29$, 6/31, 9/23, an 11/39 
(all states with question marks in Fig.~\ref{fig1}), at least in finite 
systems with appropriate electron number (see next section).

\section{Predictions for infinite systems}
\begin{figure}[t]
\epsfxsize=3.35in
\epsffile{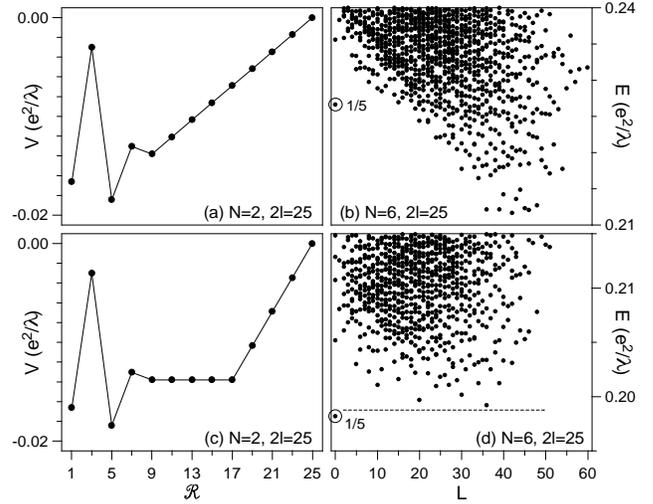}
\caption{
   Left: 
   Model pseudopotentials $V({\cal R})$, equal to $V_{\rm QE}$ in 
   Fig.~\protect\ref{fig2}a ($N=11$) for ${\cal R}\le9$ and increasing 
   to zero for ${\cal R}$ increasing to 25.
   Right: 
   Energies $E$ of six particles each with single particle energy
   $\varepsilon_{\rm QE}$ (for $\nu=1/3$ state of $N=11$ electrons)
   and interacting through pseudopotentials on the left, as a function 
   of total angular momentum $L$.   
}
\label{fig6}
\end{figure}

The extrapolation of our numerical results in order to predict stability 
of different hierarchical states in infinite systems must be done very 
carefully.
The calculations\cite{yi} show that the interaction of QP's in a Laughlin 
parent state is not generally repulsive.
The pseudopotential $V_{\rm QP}({\cal R})$ is obtained by subtraction 
of two appropriate QP energies $2\varepsilon_{\rm QP}$ from energies 
$E({\cal R})$ in Figs.~\ref{fig2} and \ref{fig3}.
E.g., for the $\nu=1/3$ parent state, the only positive pseudopotential
parameter is $V_{\rm QH}(1)$, and all others ($V_{\rm QH}({\cal R})$ for 
${\cal R}\ge3$ and $V_{\rm QE}({\cal R})$ for all ${\cal R}$) are negative.
Since $V_{\rm QP}({\cal R})$ at large ${\cal R}$ (large distance) is 
expected to vanish, it must also increase above all the values at small 
${\cal R}$ except for $V_{\rm QH}(1)$.
This brings out the question if our prediction of incompressible states 
at e.g. $\nu=6/17$ ($\nu_{\rm QE}=1/5$) or $\nu=6/19$ ($\nu_{\rm QH}=1/7$), 
verified numerically for twelve electrons (Fig.~\ref{fig5}ef) remains 
valid for an infinite system.
We have calculated energy spectra of six particles at filling $\nu=1/5$ 
($2l=25$), interacting through pseudopotentials which for ${\cal R}\le9$ 
were equal to $V_{\rm QE}({\cal R})$ for $N=11$ in Fig.~\ref{fig2}a, 
vanished at ${\cal R}=25$, and behaved in a number of different ways for 
$9<{\cal R}<25$.
As shown in Fig.~\ref{fig6}, the $L=0$ hierarchy state is always the 
lowest energy state in the low $L$ part of the spectrum, but whether
it is an absolute GS depends on how quickly $V_{\rm QP}({\cal R})$ goes 
to zero with increasing ${\cal R}$.
This is a consequence of the following operator identity
\cite{wojs-parent2,wojs-parent3} 
\begin{equation}
   \sum_{i<j} \hat{L}_{ij}^2 = \hat{L}^2 + N(N-2)\;\hat{l}^2,
\label{eqthr}
\end{equation}
which relates total ($L$) and pair ($\hat{L}_{ij}$) angular momenta
of a system of $N$ particles in a shell of angular momentum $l$.
The states with larger $L$ have (on the average) larger values of $L_{ij}$ 
and thus, if $V({\cal R})$ increases with increasing ${\cal R}$, lower 
energy.

The exact numerical calculation of $V_{\rm QP}({\cal R})$ for 
${\cal R}\gg9$, i.e. for $l_{\rm QP}\gg5$, seems impossible (calculation 
for $l_{\rm QE}=5$, i.e. $N=11$, already required diagonalization of 
a matrix with dimension nearly $10^6$), and thus the only valid test of 
stability of states like $\nu=6/17$ or 6/19 might be experiment.
However, since $V_{\rm QP}({\cal R})$ for large (infinite) system should 
be virtually zero for all ${\cal R}$'s above certain critical value, it 
is unlikely that these GS will be incompressible in the thermodynamic 
limit.
This would explain why (to our knowledge), no such states have been 
observed.

The above arguments most likely eliminate all daughter states derived 
from Laughlin $\nu=1/m$ parents with $\nu_{\rm QE}=1/5$ and $\nu_{\rm QH}
=1/7$ as possible incompressible GS's, leaving only those with $\nu_{\rm 
QE}=1$ and $\nu_{\rm QH}=1/3$.
The latter ones are incompressible, because they either correspond 
to a filled shell ($\nu_{\rm QE}=1$) or avoid the (only) repulsive 
pseudopotential parameter $V_{\rm QH}(1)>0$ ($\nu_{\rm QH}=1/3$).
A quick look at the pseudopotentials in Fig.~\ref{fig3} is enough to 
find that the incompressible states on higher levels of hierarchy can 
only have $\nu_{\rm QE}=1$.
It is easy to check (see also Fig.~\ref{fig1}) that fractions generated
in this way belong to the Jain sequence obtained from Eq.~(\ref{eqjain})
for integer $\nu^*$.
This explains why the hierarchy fractions from outside this sequence have 
not been observed experimentally,\cite{willet,mallet,sajoto,kukushkin} 
even though numerical results for finite systems presented here might 
suggest that states like $\nu=6/17$ or 6/19 are incompressible.

Let us stress that the fact that valid hierarchy states (i.e. states 
with Laughlin like correlations of appropriate Laughlin QP's) occur only 
at the Jain sequence of filling factors obviously does not contradict 
occurrence of FQH effect at other fractions.
However, the correlations in other possible FQH states, the origin 
of their incompressibility, and their elementary excitations must be 
different.
Examples of observed\cite{willet} non-hierarchy FQH states include ones 
at $\nu=7/3$ (calculations show\cite{wojs-parent3,macdonald} that it is 
not a Laughlin like $\nu=1/3$ state in the first excited LL) and $\nu=5/2$.

\section{Comparison with CF picture}
The mean field CF picture correctly predicts not only the incompressible 
states at the Jain sequence of filling factors, but also the low lying 
bands of states at any value of $N$ and $2S$.
However, neither its original justification based on cancellation 
between Coulomb and CS interactions beyond the mean field nor the use 
of higher LL's in construction of CF wavefunctions can be accepted as 
complete understanding of this success.

If the effective CF magnetic field is non-negative ($2S^*\ge0$) and the 
effective CF filling factor is less or equal than one ($\nu^*\le1$) the 
CF picture selects out of the total Hilbert space ${\cal H}$ the subspace 
${\cal H}_p$, where $2p$ is the number of bound (attached) flux quanta.
\cite{wojs-parent1,wojs-parent2,wojs-parent3} 
From Eq.~(\ref{eqcont}) and Fig.~\ref{fig1} it is clear that entire 
sequence of Jain fractions corresponds to valid hierarchy states, obtained 
by for the QE filling of $\nu_{\rm QE}=0$ or 1 on any level of hierarchy 
and/or the QH filling of $\nu_{\rm QH}=1/3$ on the first level (QH's in 
the principal Laughlin $\nu=1/m$ state).
Note that CF states obtained with $2p$ bound flux quanta for which $2S^*$ 
is positive are derived from the $\nu_{\rm QE}=1$ daughter of the Laughlin 
$\nu=(2p+1)^{-1}$ state, while those for which $2S^*$ is negative come from 
the $\nu_{\rm QH}=1/3$ daughter of the Laughlin $\nu=(2p-1)^{-1}$ state.
The explicit hierarchy wavefunctions can also be constructed
\cite{macdonald-wave,yang-wave} without introducing higher LL's 
excitations.
As demonstrated for few electron systems,\cite{batista-wave} the 
(valid) hierarchy and CF wavefunctions are nearly identical.
Another qualitative success of the CF picture, the description of higher 
bands in the energy spectrum in terms of excitations between CF LL's, in 
the hierarchy picture corresponds to creation of additional QE-QH pairs 
in the parent state.\cite{sitko-hierarchy}

One of the main results used as a direct experimental evidence for the 
existence of CF's -- the observation\cite{willet-cycl,kang-cycl} of 
geometric resonances and divergence of the CF cyclotron radius at 
$B^*\rightarrow0$ -- does not contradict the hierarchy picture where 
the fractionally charged\cite{clark} relevant QP's move in a bare 
external field $B$ so that the cyclotron radius coincides with the 
one of CF's moving in an effective field $B^*$.

\begin{figure}[t]
\epsfxsize=3.35in
\epsffile{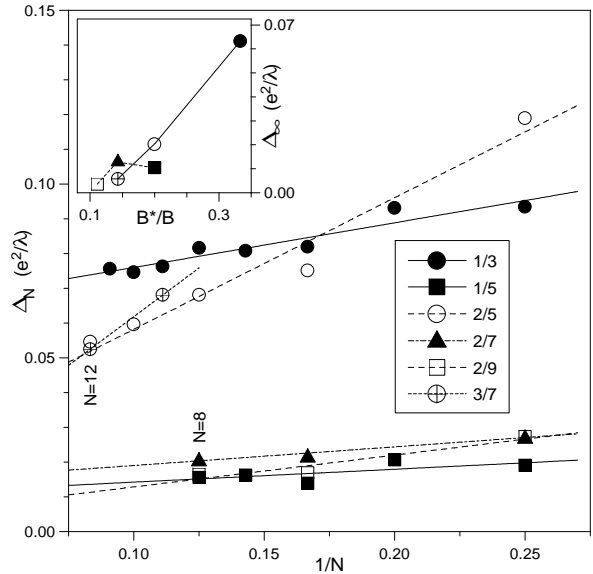}
\caption{
   Excitation gaps $\Delta_N$ of a few valid hierarchy states calculated 
   in numerical diagonalization of $N$ electrons and plotted as a function 
   of $1/N$. Inset: limiting values $\Delta_\infty$ as a function of 
   effective composite Fermion magnetic field $B^*$.
}
\label{fig7}
\end{figure}

Another result\cite{du} seemingly proving the formation of CF LL's is the 
linear dependence of the excitation gap of Jain states on the effective 
magnetic field $B^*$.
In Fig.~\ref{fig7} we plot the gaps $\Delta_N$ calculated numerically for 
a few most prominent hierarchy/Jain states as a function of the inverse 
electron number, $1/N$.
The limiting values $\Delta_\infty$ are plotted in the inset as a function 
of $B^*/B=1-2p\nu$.
The gaps of states obtained from Eq.~(\ref{eqjain}) for $p=1$ ($\nu=1/3$, 
2/5, 3/7, \dots) fall on a straight line vs. $B^*/B$ (as first observed 
by Du et al.\cite{du}; note however that the linear extrapolation to 
$B^*=0$ gives negative gaps, also in agreement with experiment of Du at 
al.\cite{du}).
However, it is not so for Jain $p=2$ states at $\nu=1/5$, 2/7, and 2/9.
In particular, the gap of the $\nu=2/7$ state seems to be larger than that 
of the $\nu=1/5$ state.
While this result may be difficult to accept in the CF picture, it is 
by no means surprising in the hierarchy picture where the relevant QP's 
in the two states are different and interact through different 
pseudopotentials (see Figs.~\ref{fig1}, \ref{fig2}, and \ref{fig3}).

The mean field CF picture and the present hierarchy picture are equivalent 
because they both use correct degeneracy of excitations, $g_{\rm QE}$ and 
$g_{\rm QH}$, for the chosen (fermionic) statistics.
The CF picture makes no use of the form of single particle wavefunctions 
in excited LL's; an effective magnetic field is just another way to obtain 
correct $g_{\rm QE}$ and $g_{\rm QH}$.
However, the authors believe that the understanding of incompressible 
states at Jain filling factors in terms of hierarchy of Laughlin 
excitations and involved QP pseudopotentials has a number of advantages 
over the CF picture.
It does not use such puzzling concepts as flux binding, depends explicitly 
and in a known way on the form of electron-electron interaction which 
enables predicting its applicability to systems with modified interactions 
(higher LL's, finite well width, etc.), and predicts correct energy gaps 
in terms of interaction parameters rather than arbitrary effective 
cyclotron energy.
The hierarchy picture is also more physically intuitive and makes the 
origin of incompressibility of Jain states more clear.
Moreover, it gives better understanding of why no other fractions are
experimentally observed, even though some (e.g. $\nu=6/17$ and 6/19)
are found in finite size numerical calculations.

\section{Conclusion}
We have calculated pseudopotentials of the interaction between 
quasiparticles, arising in the hierarchy picture of incompressible
FQH states.
Based on the analysis of these pseudopotentials, it is explained why 
no hierarchy states with filling factors $\nu$ from outside the Jain 
sequence occur in an infinite system, and thus why none have been 
observed experimentally.
Compressibility of $\nu=4/11$ and 4/13 states is demonstrated.
Laughlin like hierarchy states other than Jain states are found at 
$\nu=6/17$ and 6/19 in finite size numerical calculations.
However, it is argued that the system at these fractions (and other 
non-Jain fractions with question marks in Fig.~\ref{fig1}) will undergo 
a transition into a compressible phase when its size is increased.

The descriptions of FQH states in terms of mean field CF's and hierarchy 
of Laughlin excitations are compared.
It is explained why, despite no rigorous justification of the CF assumption 
of flux (or vortex) binding, the CF predictions and the valid predictions 
of the hierarchy picture are (qualitatively) equivalent.

In our analysis we used a fermionic statistics of quasiparticles.
However, our results are independent of this statistics and remain valid 
for hierarchy pictures formulated in terms of bosons or anyons.


\begin{references}
\bibitem{delta}
   Excitations other than containing additional QE-QH pairs are possible,
   but they have higher energy.
\bibitem{ram}
   Note that the present definition of ${\cal R}$ is slightly different 
   from that in Refs.~\onlinecite{wojs-parent1,wojs-parent3}.
\end{references}
\end{document}